*Invited talk at EUCAS'01*

Niobium based intermetallics as a source of high-current/high magnetic field superconductors

B.A.Glowacki [a,b], X-Y. Yan [a], D.Fray [a], G.Chen [a], M.Majoros [a,b] and Y.Shi [b]

[a] Department of Materials Science and Metallurgy, University of Cambridge, Pembroke Street, Cambridge CB2 3QZ, UK,

[b] Interdisciplinary Research Centre in Superconductivity, University of Cambridge, Madingley Road, Cambridge CB3 OHE, UK

*Abstract*

The article is focused on low temperature intermetallic A15 superconducting wires development for Nuclear Magnetic Resonance, NMR, and Nuclear Magnetic Imaging, MRI, magnets and also on cryogen-free magnets. There are many other applications which would benefit from new development such as future Large Hadron Collider to be built from A15 intermetallic conductors. This paper highlights the current status of development of the niobium based intermetallics with special attention to $Nb_3(Al_{1-x},Ge_x)$. Discussion is focused on the materials science aspects of conductor manufacture, such as -phase (A15) formation, with particular emphasis on the maximisation of the superconducting parameters, such as critical current density, $J_c$, critical temperature, $T_c$, and upper critical field, $H_{c2}$. Many successful manufacturing techniques of the potential niobium-aluminide intermetallic superconducting conductors, such as solid-state processing, liquid-solid processing, rapid heating/cooling processes, are described, compared and assessed. Special emphasis has been laid on conditions under which the $J_c(B)$ peak effect occurs in some of the $Nb_3(Al,Ge)$ wires.

A novel electrodeoxidizing method developed in Cambridge whereby the alloys and intermetallics are produced cheaply making all superconducting electromagnetic devices, using low cost LTCs, more cost effective is presented. This new technique has potential to revolutionise the existing superconducting industry enabling reduction of cost orders of magnitude.

**1. Introduction**

The Niobium based A-15-type superconductors that have a high critical current density, $J_c$, at very high magnetic fields are much sought after. Difficulties arise in processing a microstructure for the conductor that will not cause $J_c$ to fall off at these high fields. Since A-15 conductors are untextured in general one may assume that pinning is induced by the high angle grain boundaries and defects which are comparable to the coherence length of the particular conductor therefore $J_c$~1/grain size. This article is a review of the main process methods available in order to assess from which fabrication method the required high field properties of the most advanced A-15 conductors will be achieved.

Second generation NMR systems that can operate above 1GHz are required by industry. Only a type II A15 superconductor would currently provide any hope of this being achieved [1]. Tests were concentrated on the $Nb_3Sn$ superconductor until the 1970s, whereupon niobium aluminide ($Nb_3Al$) and $Nb_3(Al,Ge)$ was found to have potential with regards to high field, large scale applications. It is thought that using advanced $Nb_3Al$ based conductors as the inner core of the NMR machine will allow a 1GHz frequency to be achieved [2]. Unfortunately, a commercial fabrication process has not yet been established, unlike the 'bronze process' and 'copper processes' as used for its $Nb_3Sn$ counterpart [3,4]. Possibilities are severely restricted due to the difficulty in providing a sufficient value of critical current density, $J_c$, in the high magnetic field region of around 25-30 T. Through laboratory tests and theoretical knowledge, it appears that there are two main options to try and maintain a high $J_c$. One is to fabricate a β-phase $Nb_3Al$ structure, known as an A-15 phase, and through optimising the conditions try and create the highest possible $J_c$ value at high fields. The other option is to alloy the niobium aluminide with another element. This element is germanium, which brings about pinning defects in the lattice, from this a $J_c$ peak in higher fields may be observed.

## 2. $Nb_3Al$ conductors

Superconductors are defined by $J_c$, the critical temperature, $T_c$, and the upper critical field, $H_{c2}$. The $H_{c2}$ and $T_c$ are usually treated as characteristic material parameters, and are less sensitive to the microstructure, unlike $J_c$, which depends strongly on the microstructure. The processing of $Nb_3Al$ conductors can be done from one of three groups: low-temperature, high temperature or transformation processing. The A15 $Nb_3Al$ strand is processed by first making the final-size strand, with the constituents subdivided, and then heat-treating into the A15 phase.

*2.1 Low Temperature Processing*
Low temperature ( 1000°C) processes ensure that the grain size of $Nb_3Al$ does not become too coarse because the Nb/Al constituents directly react with diffusion to suppress $Nb_3Al$ grain growth. But, at low temperatures, there is a deviation from A15 stoichiometry, thus affecting high-field properties, especially $J_c$ [5].

• *Jelly-roll* (JR)- Alternate foils of Nb and Al are wound onto a copper rod and inserted into holes drilled in a copper matrix.

• *Rod-in-tube,* (RIT)- An Al-alloy rod was inserted into a Nb tube and drawn down. A triple stacking operation gives the desired 100nm core diameter to match that of the Al layer.

• *Clad chip extrusion,* (CCE)- A three-layered clad foil, Al/Nb/Al is cut into square chips, and then filled into a can in order to be extruded.

• *Powder metallurgy,* (PM)- A mixture of hydride-dehydride Nb powder and Al powder are put in a copper tube can so that they can be extruded and drawn into a monofilament wire. A bundle of these wires will give an Al layer thickness of 100nm [6].

Despite differences in cross sectional area, $J_c$ versus magnetic field curves are very similar for all processes. JR may have a slight advantage for producing long piece strands [5].

*2.2 High Temperature Processing*

At low temperatures the Nb$_3$Al phase will not be completely stoichiometric. However, at high temperatures ( 1800°C), a diffusion reaction of Nb/Al composites with laser or electron-beam irradiation allows stoichiometric A15 phase formation. Annealing at ~700°C will give better long range order, and so better $T_c$ and $H_{c2}$. Unfortunately, high temperatures will cause very coarse grains in the conductor, thereby destroying low field properties [5].

*2.3 Transformation Processing*

This process, discovered in the last few years by Kosuge et al [7], involves a combination of rapid quenching and annealing. The Nb/Al composite is quenched from 1900°C to form a bcc supersaturated solid solution of Nb(Al)ss (solid solution) and then transformation annealed below 1000°C. This process will produce Nb$_3$Al that is highly stoichiometric and has a fine grain structure, and therefore the $J_c$ will be high at both low and high fields [8]. The most common method of transformation processing is known as the rapid-heating, quenching and transformation method (RHQT), which produces lengths of conductors a few hundred metres long [7,9].

## 3. Optimising Nb$_3$Al high field properties

The main objective is to try and obtain the highest possible value for $J_c$ at high fields. However, a plot of $J_c$ versus magnetic field, fig.1, shows that a curve for a Nb$_3$Al conductor (as discussed above) will fall off quite consistently with increasing field, but for a Nb$_3$(Al,Ge) alloy, as discussed in the next section a peak is apparent just before $H_{c2}$. At present there is great debate as to which type of superconductor will ultimately give the best high field properties. It is obvious that the only way to produce a Nb$_3$Al superconductor that is practical at low and high fields is to use a transformation process. Until recently, efforts were concentrated solely on the RHQT method. However, in 2000, Takeuchi et al. [18] observed that the Nb(Al)ss obtained by RHQT from the Nb/Al composite showed a crystal structure change from a disordered to ordered structure before transforming to A15 Nb$_3$Al. This ordering appeared to be responsible for A15 phase stacking faults that reduce $T_c$, $H_{c2}$ and therefore $J_c$ at high fields. To counter this they heat treated the conductor at 1000°C, ~200°C greater than the RHQT method, which reduced the ordering and yielded a new phenomenon which they called the "transformation-heat-based up-quenching" (TRUQ). TRUQ is characterised by the self-heating of the bcc Nb(Al)ss phase by the heat of transformation. The heat propagates through the whole of the composite wire and transforms it to Nb$_3$Al [18]. Unfortunately, at this time very few papers have been published or are accessible on the subject.

*3.1 Alloying with Nb$_3$Al*

The substitution of Al with other elements such as Ge, Ga, Be, B and Cu will lead to an increase of the critical temperature, $T_c$, of the Nb$_3$Al compound. Germanium and silicon are especially good at stabilising the A15 phase present, and so recent efforts have been concentrated on the addition of Ge. At some point there must be a maximum $T_c$ for the Nb$_3$(Al,Ge) whilst still retaining its superconducting ability [4].

The most popular way to fabricate Nb$_3$(Al,Ge) is by powder metallurgy, with Nb, Al, and Ge powders encased in a Nb sheath [2]. The composite is then heat treated by laser or electron beam irradiation in order to prevent the compounds rapid grain growth. However, the cooling rate is not high enough, so $J_c$ in low fields is degraded. The most promising fabrication method is from Kikuchi et al, [19] where a rod-in-tube processed Nb/Al-20%Ge composite was rapidly heated and then quenched to form Nb$_3$(Al,Ge) filaments in a Nb matrix, from which an A15 type phase was formed directly. Heat treatment at 800°C for 10 hours will improve the A15 long range order, and therefore $T_c$. The $J_c$ value at high fields is now much higher than for RHQT Nb$_3$Al. Further refinement is required to reduce the Nb matrix ratio without affecting the mechanical properties.

*3.2 Improving the $J_c$ peak at high fields*

In order to improve the usefulness of superconducting conductor for the NMR magnet, the peak must move further to the higher magnetic field values of a $J_c$ versus $H$ plot (see fig. 1). Magnetic pinning interactions occurring at boundaries of different phases (which will have different magnetisation curves) will bring about minimum point for $J_c$ versus $H$ where the two phases have equal magnetisation values and that $J_c$ and $F_p$ (the pinning force) will therefore be zero. So at some point between minimum and $H_{c2}$ there will be a $J_c$ peak, called a 'valley effect' as noted by Evetts [20]. Magnetic pinning is not the only type of mechanism that results in peaks, there is also a mechanism brought about by flux lattice defects and weak 'line-like' pinning forces as proposed by Pippard. Pippard observed that the $J_c$ peak would always occur near to $H_{c2}$. His explanation was that the rigidity of the flux lattice falls to zero at $H_{c2}$ more rapidly than the pinning force of the inhomogeneities. This loss of rigidity would strengthen the flux against unpinning by forces due to the current because the flux lattice will more readily form a "minimum energy flux pattern" [21].

Osborne and Kramer looked further into the effect of plastic deformation and pinning by plotting pinning force per unit volume against a reduced magnetic field, $h = H/H_{c2}$. A peak was observed for wider conditions than if $J_c$ were against $H$. The $J_c$ peak is then a development of a narrow peak in the pinning force at high-reduced fields. Microstructures that give a high density of weak pins, or moderate density pins will give the required narrow peak of $F_p$ at high h. It was also found that increasing plastic deformation, and therefore the pin density, will shift the $F_p$ peak to lower $h$, broaden the peak and increase the height without increasing $F_p$ at high $h$. From this it is obvious that fabrication methods must not involve too much deformation [22]. A high density of defects in a flux-line lattice (FLL) means that fluxoids will fit more readily to the pinning centres, thus increasing $J_c$ [23]. So, for Nb$_3$Al and some other superconductors, an increasing $J_c$ with increasing magnetic field below the peak is considered to result from a deformation of the flux-line lattice with defects to fit the pinning centres more easily. This occurs until depinning happens, whereupon $J_c$ dramatically reduces [2].

Maximisation of the $J_c$ peak and its position for Nb$_3$(Al,Ge) is difficult because the sensitivity of intermetallic formation process to thermal and mechanical processing conditions. At present, it is widely accepted that Nb$_3$(Al,Ge) – although what quantity of Ge is unclear - wires should be made by the rapid-heating/quenching process because the TRUQ method has not been developed enough. Recent experiments have found that $J_c$ at high fields will increase with decreasing Al-Ge alloy core size in the precursor wire. For example, a 0.3 micron-core Nb$_3$(Al,Ge) wire has a $J_c$ (4.2K, 25T)  150 A/mm$^2$ which is almost twice as large as a 1.5 µm core. This is due to the reduction of Al-Ge alloy diameter enhancing the volume fraction of the A15- Nb$_3$(Al,Ge) phases by increasing the diffusion pair density in the composite [19].

## 4. Direct electrochemical reduction of Nb-based oxides DERO (the method to produce low cost multifilamentary superconducting conductors)

The recently introduced process for extracting metals and alloys from solid oxides by direct electrochemical reduction in molten salt (DERO) known also as a Fray-Farthing-Chen

Cambridge Process, FFCCP, is expected to revolutionise the existing superconducting manufacture procedures, enabling reduction of cost an order of magnitude [24]. In this particular case the complete conductor manufacture procedure has to have four distinctive processing stages: 1) electrochemical reduction of the Nb-based compounds, 2) infiltration by Al-based alloys (or any other elements or alloys to form intermetallics or the artificial pinning centres), 3) deformation processes, and 4) reactive formation of the intermetallic layer followed by insulation process.

*4.1 Direct Electrochemical Reduction of Oxides (DERO)*
The electrochemical reduction route is a much easier, quicker and cheaper way to extract many metals and alloys than the established metallurgical routes. The schematic of such a process is presented in fig.2.

In the case of the simple binary alloy such as NbTi, the preforms of mixed Nb-O, Ti-O powders can be made as the cathode in molten $CaCl_2$ whose cation can form a more stable oxide, CaO. The oxygen in Nb-Ti-O is simply ionised and dissolves in the salt, leaving Niobium-Titanium alloy metal of the desired composition behind. The extraction of the Nb, NbTi and $Nb_3Sn$ (fig.3) metals from oxides using the process was already demonstrated on the laboratory scale.

*4.2 Infiltration of NbTi and A-15*
After electrochemical reduction and cleaning, the final product of process is a porous metallic sponge. The degree of porosity of the final percolative network of Nb-based alloy will strongly depend on the density of the oxide preform and also on an initial preparation technique of the prefabricated oxide. The sintered pellets show a significant shrinkage and greatly increased strength in comparison with those prepared from the powders. Because the resulting metallic product of the electrochemical reduction is soft and porous, without structural defects and secondary phases, it can not be regarded as a high critical current density, $J_c$, superconducting material.

Such semi-finish product has to be upgraded by introduction of Al or Sn alloys for the reactive diffusion formation of the intermetallic phase in the case of A15 intermetallic superconductors, or/and by introduction of the Artificial Pinning Centres as in the case of NbTi. However attractive, all three existing APC techniques require very expensive metal alloy rods and extensive cold working processes from very large diameters [25-27].

Infiltration of the presintered alloy rod or plate such as Nb-Ti-X or Nb-Al-Ge-Z by Al [28] or Al alloy material, where X or Z represents addition metals has its advantages because for example recently NbTiTa ternary alloys have been explored as a high field avenue for APC materials.

After the infiltration process, rods would be machined to the desired shape and inserted in subsequent tubes serving as diffusion barrier and electrical and thermal stabilisation. Although an elevated temperature during the infiltration stage may produce some A15 phase, it will be desired to subject the infiltrated tape to a substantial reduction in thickness by cold rolling prior to the final diffusion formation of the intermetallic A15 layers in the conductor.

## 5. Conclusions

This review gives a summary of the developments of $Nb_3Al$ superconductors during the last 30 years. Whilst by no means fully comprehensive, it does give some ideas required to maximise $J_c$ at high fields. Although processing $Nb_3Al$ conductors has recently seen a resurgence due to new techniques, it seems clear that the $Nb_3(Al,Ge)$ alloys and their $J_c$ peaks will lead the way to establishing a second generation NMR systems that can operate above a frequency of 1GHz.

Proved new, the low cost DERO process for reducing solid oxides to metals and alloys of the pre-defined alloy composition in our opinion opens new opportunities for the manufacture of the highest quality low temperature superconductors. This process is applicable not only to the most saleable ones such as NbTi but also those best intermetallic ones which are difficult to manufacture such as $(Nb,X)_3(Sn,Z)$ and $Nb_3(Al,Ge)$ characterised by the highest $J_c$, $B_{c2}$, $T_c$ values. Novel conductors will be produced for the fraction of the cost of the currently available conductors. This is a very exciting prospect which, on present evidence, shows every chance of being realised.

The DERO process is going to be applied to the niobium-titanium, niobium-tin and niobium-aluminium systems because of the supportive research attention it would receive in the professional world and literature and should speed up the development of the practical conductors.

**Figure captions:**

Fig. 1   Critical current versus external magnetic field for the $Nb_3Al$ and $Nb_3(Al,Ge)$ conductors manufactured by different processes as specified in the legend. For $Nb_3(Al,Ge)$ tapes formed by -phase and liquid quenching processes magnetic field was parallel to the tape surface. In all cases magnetic field was perpendicular to the current.

Fig. 2   Schematic of the electrolytic cell for the reduction of Nb-O and Ti-O bar suspended in the molten $CaCl_2$ salt.

Fig. 3   AC susceptibility of the Nb, NbTi and $Nb_3Sn$ metallic rods manufactured by DERO process. Reduction of the identical $Nb_2O_5$-$SnO_2$ samples to $Nb_3Sn$ was conducted under identical voltage and current conditions but at different reduction temperature $T_r$ in the molten $CaCl_2$ bath. The values of the $T_c$ Nb, NbTi and $Nb_3Sn$ onset are in a very good agreement with the standard material data. There is some formation of the  Nb-Sn manifested by reduction of $T_c$ value to 6.5K.

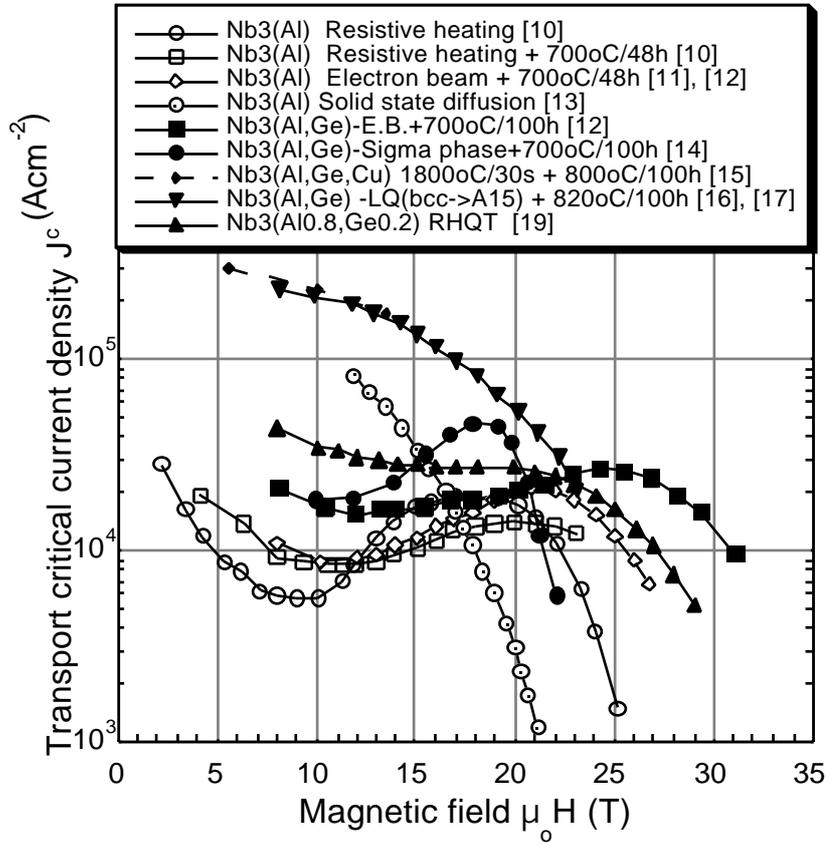

Figure 1

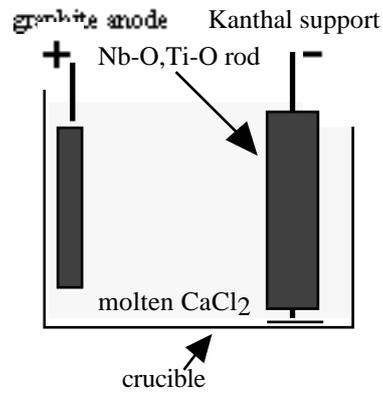

Figure 2

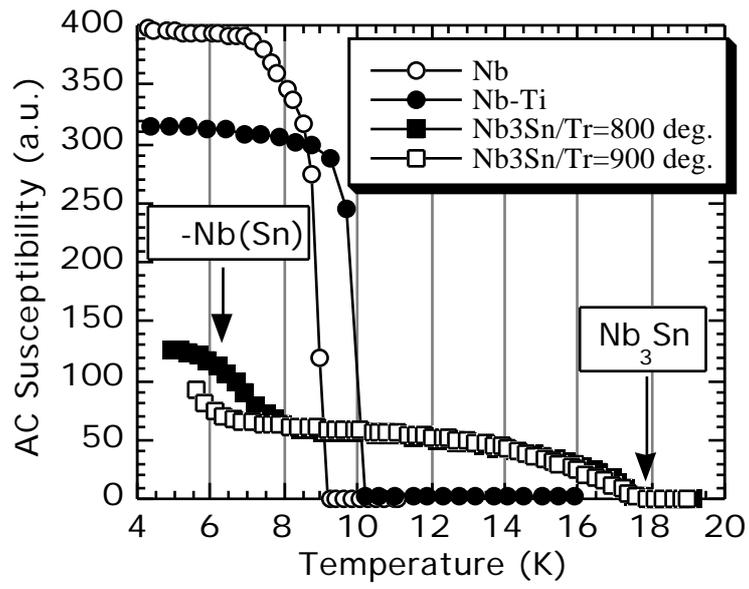

Figure 3